\documentclass[twocolumn,prl]{revtex4}
\usepackage{amssymb}
\usepackage{amsthm}
\usepackage{color}
\usepackage{rsfs}
\usepackage{ulem}
\usepackage[all]{xy}
\usepackage{amsmath}
\usepackage{graphicx}
\newtheorem*{theorem*}{Theorem}
\newtheorem*{definition*}{Definition}
\newcommand{\be}{\begin{equation}}
\newcommand{\ee}{\end{equation}}
\newcommand{\bea}{\begin{eqnarray}}
\newcommand{\eea}{\end{eqnarray}}

\newcommand{\ra}{\rangle}
\newcommand{\pa}{\partial}
\newcommand{\la}{\langle}

\begin{document}

% Use the \preprint command to place your local institutional report
% number in the upper righthand corner of the title page in preprint mode.
% Multiple \preprint commands are allowed.
% Use the 'preprintnumbers' class option to override journal defaults
% to display numbers if necessary
%\preprint{}

\title{Holographic Interpretation of Shannon Entropy of Coherence of Quantum Pure States}

% repeat the \author .. \affiliation  etc. as needed
% \email, \thanks, \homepage, \altaffiliation all apply to the current
% author. Explanatory text should go in the []'s, actual e-mail
% address or url should go in the {}'s for \email and \homepage.
% Please use the appropriate macro foreach each type of information

% \affiliation command applies to all authors since the last
% \affiliation command. The \affiliation command should follow the
% other information
% \affiliation can be followed by \email, \homepage, \thanks as well.
\author{Eiji Konishi\footnote{E-mail address: konishi.eiji.27c@kyoto-u.jp}}

%\email[]{}
%\homepage[]{Your web page}
%\thanks{}
%\altaffiliation{}

\affiliation{Graduate School of Human and Environmental Studies, Kyoto University, Kyoto 606-8501, Japan}

%Collaboration name if desired (requires use of superscriptaddress
%option in \documentclass). \noaffiliation is required (may also be
%used with the \author command).
%\collaboration can be followed by \email, \homepage, \thanks as well.
%\collaboration{}
%\noaffiliation

\date{\today}

\begin{abstract}
For a quantum pure state in conformal field theory, we generate the Shannon entropy of its coherence, that is, the von Neumann entropy obtained by introducing quantum measurement errors.
We give a holographic interpretation of this Shannon entropy, based on Swingle's interpretation of anti-de Sitter space/conformal field theory (AdS/CFT) correspondence in the context of AdS$_3$/CFT$_2$.
As a result of this interpretation, we conjecture a differential geometrical formula for the Shannon entropy of the coherence of a quantum pure or purified state in CFT$_2$ at thermal and momentum equilibrium as the sum of the holographic complexity and the abbreviated action, divided by $\pi\hbar$, in the bulk domain enclosed by the Ryu--Takayanagi curve.
This result offers a definition of the action of a bulk model of qubits dual to the boundary CFT$_2$ at this equilibrium.
\end{abstract}


%\maketitle must follow title, authors, abstract, \pacs, and \keywords
\maketitle
% body of Letter here - Use proper section commands
% References should be done using the \cite, \ref, and \label commands

% ----------------------------------------------------------
% INTRODUCTION
% ----------------------------------------------------------

% ----------------------------------------------------------
%  von Neumann's two ideas of quantum measurement
% ----------------------------------------------------------

{\it Introduction.---}In Ref.\cite{vN}, von Neumann proposed basic ideas for two methods of quantum measurements.
The first method is to trace out the quantum state of the measurement apparatus, that is, the environment of the measured system after the measurement apparatus and the measured system interact and their states become entangled.
The second method is the use of the superselection rules obtained by introducing quantum measurement errors.\cite{MS2,MS1a,MS1b}
These quantum measurements are changes of the given pure state of the measured system to a mixed state, that is, an exact statistical mixture of pure eigenstates of the measured quantity.
So, we can identify these quantum measurements with the sources that generate the von Neumann entropy.

% ----------------------------------------------------------
%  Entanglement entropy
% ----------------------------------------------------------

Recently, the translation of the von Neumann entropy, which accompanies the change of a quantum pure state to a statistical mixture, into a differential geometrical quantity in the language of gravity via anti-de Sitter space/conformal field theory (AdS/CFT) correspondence\cite{AdSCFT1,GKP,W,AdSCFT2}---the example of the holographic principle\cite{Hol1,Hol2,Hol3}---has been extensively studied.\cite{Tak}
The renowned Ryu--Takayanagi formula and its covariant extension\cite{RT1,RT2,HRT,Review} apply von Neumann's first method of quantum measurement to the pure state of two mutually entangled quantum systems that are geometrically complementary to each other in CFT and interpret the generated von Neumann entropy, called the {\it entanglement entropy}, of the target system as a differential geometrical quantity on the gravity side.

% ----------------------------------------------------------
%  Shannon entropy
% ----------------------------------------------------------

However, no study has addressed the holographic dual of the von Neumann entropy  generated by von Neumann's second method of quantum measurement that can be applied to an arbitrary quantum pure state of the target system.

% ----------------------------------------------------------
%  In this Letter, ...
% ----------------------------------------------------------

In this paper, we attempt to translate into the language of gravity the Shannon entropy of the coherence of a quantum pure or purified state in CFT$_2$ without gravity at thermal and momentum equilibrium; this Shannon entropy is obtained by the second method of quantum measurement and is the information lost by describing the quantum pure state in the classical way.
This translation is done by using AdS$_3$/CFT$_2$ correspondence based on Swingle's interpretation\cite{AdSCFT1,GKP,W,AdSCFT2,Swingle} and provides a definition of the action of a bulk model of qubits dual to the boundary CFT$_2$ at this equilibrium.
Throughout this paper, we choose $(-,+,+)$ as the signature of the AdS$_3$ space-time metric and adopt the natural system of units, and hatted variables are quantum operators.

\smallskip
\smallskip

{\it Generation of the von Neumann entropy.---}From here, we study the holographic dual of the origin of the von Neumann entropy generated by the second method of quantum measurement of a quantum system $A$.

For the sake of simplicity, $A$ consists of a particle up to (\ref{eq:end}).
The argument is based on the mathematical theorem below, which was proven by von Neumann.\cite{vN,MS2}
\begin{theorem*}[von Neumann]
Consider the measurement errors $\sigma_Q$ and $\sigma_P$ of the canonical variables $\widehat{Q}$ (position) and $\widehat{P}$ (momentum), respectively, of a spatial dimension of $A$.
Then, under the condition
\begin{equation}
\sigma_Q\sigma_P < 60^2 \frac{\hbar}{2}\;,\label{eq:error}
\end{equation}
there are errors $\sigma_Q$ and $\sigma_P$ such that the redefined $\widehat{Q}$ and $\widehat{P}$ that incorporate these errors commute with each other and are able to be simultaneously measured.
\end{theorem*}
In this theorem, we partition the spectrum lines of $\widehat{Q}$ and $\widehat{P}$ into an infinite number of intervals of constant widths $\epsilon_Q$ and $\epsilon_P$ (which satisfy the equality $\epsilon_Q\epsilon_P=h$\cite{vN}), respectively, and we replace these canonical variables with $\widehat{Q}^\prime$ and $\widehat{P}^\prime$, which take constant eigenvalues in each of their intervals.
Specifically, the spectra of $\widehat{Q}^\prime$ and $\widehat{P}^\prime$ are
\begin{eqnarray}
{\rm Spec}\widehat{Q}^\prime:0,\pm \epsilon_Q,\pm 2\epsilon_Q,\ldots\label{eq:SpecQ}
\end{eqnarray}
and
\begin{eqnarray}
{\rm Spec}\widehat{P}^\prime:0,\pm \epsilon_P,\pm 2\epsilon_P,\ldots\;,\label{eq:SpecP}
\end{eqnarray}
respectively.
In this theorem, because $\widehat{Q}^\prime$ and $\widehat{P}^\prime$ commute, we can consider the phase space of $A$ as in classical mechanics.
This phase space is divided into the Planck cells specified by (\ref{eq:SpecQ}) and (\ref{eq:SpecP}), and there is one possible redefined quantum orbital state for each Planck cell.

When all (orbital) observables $\widehat{{\cal O}}$ of $A$ are able to be simultaneously measured within certain measurement errors $\sigma_{{{{\cal O}}}}$ (i.e., when only such approximate measurements are made for $A$), we say that $A$ is in a {\it macroscopic state} (distinct from the term {\it macrostate} in statistical mechanics).\cite{MS2,MS1a,MS1b}
A macroscopic state of $A$ is determined by the corresponding set of observables $\{\widehat{{\cal O}}\}$ of $A$ that is restricted from the total set of Hermitian operators.
This set of observables is a set of commutative elements because they can be simultaneously measured (i.e., are classicalized).\cite{MS1a,MS1b}\footnote{When we introduce measurement errors, we assume that the system is in a macroscopic state.}
Here, note that all orbital observables, that is, all polynomials of the redefined position and momentum operators of the particle, are classicalized in the sense of {\it classicalization} $\grave{{\rm a}}$ la von Neumann.

% ----------------------------------------------------------
%  From pure state to mixture: generation of von Neumann entropy
% ----------------------------------------------------------

Now, in a macroscopic state of $A$, for any polynomial operator $\widehat{{\cal{O}}}$ of $\widehat{Q}^\prime$ and $\widehat{P}^\prime$, the superselection rules
\begin{equation}
[\widehat{{\cal{O}}},\widehat{Q}^\prime]=[\widehat{{\cal O}}, \widehat{P}^\prime]=0
\end{equation}
hold.
We denote by ${\cal R}^\prime$ a pair comprising the redefined position $Q^\prime$ in the spectrum (\ref{eq:SpecQ}) and the redefined momentum $P^\prime$ in the spectrum (\ref{eq:SpecP}).
When ${\cal R}^\prime_1$ and ${\cal R}^\prime_2$ are distinct, $|{\cal R}^\prime_1\ra$ and $|{\cal R}^\prime_2\ra$ have no coherence.\cite{MS2,MS1a,MS1b}
The reason for this is that
\begin{eqnarray}
[\widehat{{\cal{O}}},\widehat{{\cal R}}^\prime]=0\Rightarrow\la {\cal R}^\prime_1|\widehat{{\cal{O}}}|{\cal R}^\prime_2\ra=0\label{eq:key}
\end{eqnarray}
holds for $\widehat{{\cal R}}^\prime\equiv \widehat{Q}^\prime,\widehat{P}^\prime$.
From this fact, when $A$ is in a macroscopic state, the superposition of simultaneous eigenstates $\{|{\cal R}^\prime\ra\}$ of $\widehat{Q}^\prime$ and $\widehat{P}^\prime$, made distinct by spectrum width, becomes a statistical mixture: the von Neumann entropy is generated.
Indeed, by expanding the given quantum pure state $|\psi_A\ra$ to
\begin{equation}
|\psi_A\ra=\sum_nc_n|\psi_n^{{\rm basis}}\ra\;,\label{eq:psiA}
\end{equation}
it follows that
\begin{eqnarray}
\widehat{\varrho}_A=\sum_{m,n}c_m\bar{c}_n|{\psi}_m^{\rm basis}\ra \la {\psi}_n^{\rm basis}|\to \sum_n|c_n|^2|{\cal R}_n^\prime\ra \la {\cal R}_n^\prime|\;,\label{eq:change}
\end{eqnarray}
where $\la Q|\psi_n^{\rm basis}\ra\equiv\la Q|{\cal R}^\prime_n\ra$ are Gaussian wave packets of the original position $Q$ with width $\epsilon_Q$ and Fourier width $\epsilon_P$, and their full set forms a complete orthonormal system of the functional Hilbert space of $A$\cite{vN}.
Namely, by introducing measurement errors of $A$ that satisfy (\ref{eq:error}), the von Neumann entropy
\begin{eqnarray}
S_A&\equiv&-{\rm tr}\widehat{\varrho}_A\ln \widehat{\varrho}_A\label{eq:SA}\\
&=&-\sum_np_A[{\cal R}_n^\prime]\ln p_A[{\cal R}_n^\prime]\;,\ \ p_A[{\cal R}_n^\prime]=|c_n|^2\label{eq:Shannon}
\end{eqnarray}
is generated from the superposition of the simultaneous eigenstates $\{|{\cal R}^\prime\ra\}$ of $\widehat{Q}^\prime$ and $\widehat{P}^\prime$ (an arbitrary quantum pure state) of $A$.
By using the Shannon entropy in bits
\begin{equation}
H_A^{\rm bit}=-\sum_np_A[{\cal R}_n^\prime]\log_2p_A[{\cal R}_n^\prime]\;,\label{eq:HAbit}
\end{equation}
we can write the von Neumann entropy as
\begin{equation}
\frac{S_A}{\ln 2}=H_A^{\rm bit}\;.\label{eq:end}
\end{equation}
Equation (\ref{eq:HAbit}) is the information lost by introducing measurement errors (i.e., by the classicalization of the given quantum pure state $|\psi_A\ra$).

Finally, we consider the case of a system in quantum field theory (QFT).
For a system in QFT, we adopt the interaction picture in order to define a particle (i.e., its position and momentum eigenstates) of a field.
In the classicalization of a system in QFT$_2$, due to the measurement errors, to describe the positions and momenta of particles with sufficient fineness, we must introduce the ultraviolet cutoff, that is, the lattice constant $\epsilon$, for the boundary line and its reciprocal constant (refer to the fifth paragraph of the next section).
Then, noting that all redefined orbital number operators are classicalized (i.e., have simultaneous eigenstates), the Shannon entropy of the coherence of a system in QFT$_2$ on this lattice and its reciprocal (i.e., a quantum many-body system) is defined as in quantum mechanics as explained so far.

% ----------------------------------------------------------
% Application of AdS/CFT correspondence
% ----------------------------------------------------------

\smallskip
\smallskip

{\it AdS/CFT correspondence.---}Next, we apply the AdS$_3$/CFT$_2$ correspondence based on Swingle's interpretation to a quantum pure state $|\psi_A\ra$ of a quantum system $A$ in a strongly coupled CFT in two-dimensional Minkowski space-time.\cite{AdSCFT1,GKP,W,AdSCFT2,Swingle}
In this application, we consider the change of bulk induced by the measurement errors.
In this section, we consider static and classical bulk space-time at zero boundary temperature, which corresponds to the ground state of an isolated quantum system $A$.

In the bulk space, we consider the {\it bubble}, defined for the domain $A$ sitting on the boundary,
\begin{equation}
\gamma_A\;,\ \ {\rm s.t.}\ \ \pa\gamma_A=\pa A\;,\ \ \gamma_A\sim A\;.
\end{equation}
The symbol $\sim$ indicates homology equivalence.

In Swingle's interpretation based on Vidal's {\it multiscale entanglement renormalization ansatz} (MERA), a quantum pure state on a lattice in the boundary space is equivalent to the tensor network of qubits
(here, we assume that the fermionic occupancy/empty states at each site on the lattice in the second quantization characterize the qubit states)
obtained by the entanglement renormalization group transformation in the bulk space and each bubble $\gamma_A$ of $A$ transverses this tensor network.\cite{Swingle,Vidal,Vidal2}
Here, the {\it entanglement renormalization group transformation} is the real-space renormalization group transformation of the semi-infinitely alternate combinations of the {\it disentangler} that makes the entanglement between two adjacent qubits be the product state by a unitary transformation and an {\it isometry} that adjoins two adjacent qubits into one qubit by coarsening the grain.\cite{Vidal}
The sequence of this transformation of qubits is a scale-invariant tensor network.\cite{Vidal,Vidal2}

Now, we denote by $G_N$ the three-dimensional Newton's gravitational constant.
Then, as the Ryu--Takayanagi formula of the entanglement entropy\cite{RT1,RT2}
\begin{equation}
S_A^{\rm EE}=\min_{\gamma_A}\frac{{\rm Length}[\gamma_A]}{4G_N}\label{eq:RT}
\end{equation}
holds in a static AdS$_3$ space-time, it is known that, for a sufficiently strongly entangled system $A$,
\begin{eqnarray}
S_A^{\rm EE}\propto\min_{\gamma_A}\sharp\{{\rm EPR\ pairs\ crossed\ by}\ \gamma_A\ {\rm in\ bulk}\}\label{eq:RT2}
\end{eqnarray}
holds in Swingle's interpretation.\cite{Tak}
Here, denoting by $\gamma_A^0$ the geodesic bubble, $\gamma_A=\gamma_A^0$ gives rise to the values of (\ref{eq:RT}) and (\ref{eq:RT2}).
Before we introduce the measurement errors, the {\it holographic dual}, $\chi_A$, of the quantum pure state of $A$ in a static AdS$_3$ space-time is, by formula (\ref{eq:RT2}), the bulk domain whose boundary is specified by the geodesic bubble $\gamma_A^0$ and $A$ (the so-called {\it time slice of the entanglement wedge} of $A$ in the bulk):
\begin{equation}
\Gamma_A^0\;,\ \ \pa \Gamma_A^0=\gamma_A^0\cup A\;.\label{eq:object}
\end{equation}

Next, we introduce the quantum measurement errors $\sigma_Q$ and $\sigma_P$ fine enough to describe positions and momenta of the system $A$ such that the spectrum lattices of (\ref{eq:SpecQ}) and (\ref{eq:SpecP}), respectively, match the lattice and its reciprocal.
Then, over the whole of the boundary, no redefined position eigenstate has coherence with any other due to (\ref{eq:key}).
So, as in (\ref{eq:change}), the state of an arbitrary system of different sites on the boundary is a statistical mixture of product eigenstates of redefined positions and has no entanglement.
From this fact, in the bulk space, there are no disentangler operations in the entanglement renormalization group transformation, by which the tensor network (that is, the holographic dual of the quantum systems on the boundary) works.
Namely, both on the whole of the boundary and also in the whole of the tensor network in the bulk space, there is no coherence and there are no EPR pairs.
Consequently, for the pure states in the mixed state of $A$, all the bubbles $\gamma_A$ containing $\gamma_A^0$ and all the bulk domains $\Gamma_A$ (s.t. $\pa \Gamma_A=\gamma_A\cup A$) containing $\Gamma_A^0$, the time slice of the entanglement wedge of $A$ in the bulk, cannot be distinguished from each other.
In other words, information of $\Gamma_A^0$ as the time slice of the entanglement wedge of $A$ in the bulk is lost.

From this consequence, the holographic dual $\chi_A$ of $A$ in the bulk space has a {\it microstate} (in the sense of statistical mechanics) for each bulk domain $\Gamma_A$ and has entropy $S_A^{({\rm bulk})}$ with value (\ref{eq:Shannon}) from the holographic principle.

As statistical mechanics reveals, the origin of the bulk entropy $S_A^{({\rm bulk})}$ generated by the measurement errors is the number of microstates of $\chi_A$.
(Before we introduce measurement errors, (\ref{eq:SA}) is zero and the microstate of $\chi_A$ is uniquely specified by $\Gamma_A^0$.)
This origin is clearly different from the origin of the bulk entropy generated by the partial trace over the geometrically complementary system on the boundary (i.e., by the first method of quantum measurement).

% ----------------------------------------------------------
% Holographic conjecture
% ----------------------------------------------------------

\smallskip
\smallskip

{\it Holographic interpretation.---}In the second method of quantum measurement, the holographic interpretation follows from the above arguments.

For a given quantum pure state $|\psi_A\ra$ of the boundary system $A$ in contact with an energy and momentum reservoir\cite{McLennan,MIH}\footnote{After introducing quantum measurement errors, the equilibrium state of CFT$_2$ can be purified by means of the tilde states $|\{\widetilde{{\cal R}}^\prime\}\ra$.\cite{MIH}}, we consider the Shannon entropy $H_A^{\rm bit}$ of its coherence in bits.
This entropy is the information lost by the introduction of measurement errors.
We assume that the measurement errors $\sigma_Q$ and $\sigma_P$ are fine enough to describe $A$, as mentioned in the last section.
In our setup, we propose that
\begin{equation}
H_A^{\rm bit}=C_A+\frac{1}{\pi\hbar}I_A\;,\label{eq:result}
\end{equation}
where $C_A$ is the holographic complexity of $A$ and is given essentially by the area of the holographic dual $\Gamma_A^0$ of $A$ for the geodesic bubble $\gamma_A^0$\cite{Complexity1,Complexity2,Complexity3,Alishahiha}\footnote{So neither $C_A$ nor (\ref{eq:change}) contains information about the relative phases in $|\psi_A\ra$.}, and $I_A$ is the value of the action $I[\Gamma_A]$ taken at $\Gamma_A^0$.

Here, the action $I[\Gamma_A]$ is defined in the following way.
First, we decompose the square of line elements of the bulk space-time by using the metric $\gamma_{\mu\nu}$ of the two-dimensional space-times and an Arnowitt--Deser--Misner-like method as
\begin{equation}
ds^2=N^2dr^2+\gamma_{\mu\nu}(dx^\mu+v^\mu dr)(dx^\nu+v^\nu dr)\;,\label{eq:ds2}
\end{equation}
where $r$ is the extra dimension.\cite{AdSCFT3}
In this decomposition, the action $I[\Gamma_A]$ is defined, in the family of phase spaces of the two-dimensional space-like hypersurface $\Gamma^2$
\begin{equation}
{\mathfrak M}_r=\{(x,p^x)|\ p^x\ {\rm at}\ (x,r)\}\label{eq:Mr}
\end{equation}
for momentum density $p^x$ of the slice of $\Gamma^2$ at each position $r$ of the extra dimension\cite{AdSCFT3}, as
\begin{eqnarray}
I[\Gamma_A]\equiv \int_{\Gamma_A} p^xv_x d\Gamma_A\;,\label{eq:action}
\end{eqnarray}
where the index $x$ is contracted by the Kronecker delta (see the arguments after (\ref{eq:ds})).
This action is the abbreviated action in the family of phase spaces ${\mathfrak M}_r$.

Proposal (\ref{eq:result}) is based on the following arguments.
First, we expand the given quantum pure state $|\psi_A\ra$ of the system $A$ to $|\psi_A\ra=\sum_nc_n|\{{\psi}^{\rm basis}\}_n\ra$ and set $p_A[\{{{\psi}^{\rm basis}}\}_n]=|c_n|^2$.
Here, $|\{{{\psi}^{\rm basis}}\}_n\ra$ are equivalent to distinct product simultaneous eigenstates $|\{{\cal R}^\prime\}_n\ra$ of the redefined positions $\{\widehat{Q}^\prime\}$ and redefined momenta $\{\widehat{P}^\prime\}$ with many-body degrees of freedom; we omit the tilde part\cite{MIH}.
Next, we consider the bulk side.
After introducing quantum measurement errors, $\Gamma_A$ (that is, a microstate of $\chi_A$) has a normalized distribution function, $f[\Gamma_A]$.
As the consequence of the last section's main argument, instances of $\Gamma_A$ cannot be distinguished from each other and are uniformly distributed over their configuration space.
So, due to Tolman's {\it principle of equal a priori probabilities}\cite{Tolman},
\begin{eqnarray}
f[\Gamma_A]=\frac{1}{W_A^{({\rm bulk})}}
\end{eqnarray}
holds for the number of states $W_A^{({\rm bulk})}$.
Here, the {\it equal a priori probability} $p_A^{({\rm bulk})}\equiv 1/W_A^{({\rm bulk})}$ is determined by the horizontal bulk-space momentum density $p^x$, which is an additive quantity, its multiplier for the momentum constraint, and the geometry of $\Gamma_A^0$ via maximization of the number of microstates subject to the {\it a posteriori} number and momentum constraints in the bulk space.
In the gravitational theory dual to the CFT of the boundary, the multiplier of $p^x$ is the `{\it shift vector}' $v_x$.
This multiplier $v_x$ is obtained from the thermodynamic relation\footnote{Note that $d(1/T_{\rm bulk})\propto -dr/r$, from the arguments that follow.}
\begin{equation}
\frac{\pa \Delta s_A^{({\rm bulk})}}{\pa p^x}=-v_x\Delta \biggl(\frac{1}{T_{\rm bulk}}\biggr)
\end{equation}
for the horizontal bulk-space entropy density $s_A^{({\rm bulk})}$ (s.t. $\Delta s_A^{({\rm bulk})}=\int (dS_A^{({\rm bulk})}/\sqrt{g_{xx}})/dx$ for the area form of the bulk-space entropy density $dS_A^{(\rm bulk)}$) and the inverse bulk-space temperature $1/T_{\rm bulk}$ ($0\le 1/T_{\rm bulk}\le \infty$) under the identification
\begin{equation}
\frac{1}{T_{\rm bulk}}=-\frac{R_{\rm AdS}{\cal U}}{\kappa\hbar}\label{eq:id}
\end{equation}
for the curvature radius $R_{\rm AdS}$ of AdS$_3$ space-time, a dimensionless `{\it time}' parameter ${\cal U}$ defined by the relation $r=r_\infty 2^{\cal U}$ for the ultraviolet cut-off $r_\infty$ of the extra dimension $r$, and a dimensionless positive real constant $\kappa$: this is when we regard the entanglement renormalization group transformation as being like inverse time evolution\cite{Tak}.
(Here, the range of ${\cal U}$ is $-\infty \le {\cal U}\le 0$ at zero boundary temperature\cite{Swingle}; at finite boundary temperatures, we paste two truncated MERA networks together at the position of the black hole event horizon\cite{MIH}.)
Then, the {\it equal a priori probability} $p_A^{({\rm bulk})}$ is the value of the next factor taken at the holographic dual $\Gamma_A^0$ of the quantum pure state of $A$:
\begin{equation}
p_A^{({\rm bulk})}=\exp\biggl(-\frac{\ln 2}{8\pi G_NR_{\rm AdS}}{\rm Area}[\Gamma_A^0]-\frac{\ln 2}{\pi\hbar}I[\Gamma_A^0]\biggr)\;.\label{eq:pAbulk}
\end{equation}
This probability gives rise to the entropy $S_A^{({\rm bulk})}$ of $\chi_A$ (i.e., the information lost about $\Gamma_A^0$).\footnote{See the appendix for the derivation of (\ref{eq:pAbulk}).}

From the above arguments and the holographic principle, the Shannon entropy $(\ln 2)H_A^{\rm bit}$ of the quantum mixed state of $A$ in nats is
\begin{eqnarray}
(\ln 2)H_A^{\rm bit}&=&-\sum_np_A[\{{\psi}^{\rm basis}\}_n]\ln p_A[\{{{\psi}^{\rm basis}}\}_n]\label{eq:start}\\
&=&S_A\\
&=&S_A^{({\rm bulk})}\label{eq:hol}\\
&=&-\int f[\Gamma_A]\ln f[\Gamma_A]{\cal D}^f\Gamma_A \\
&=&\ln W_A^{({\rm bulk})} \\
&=&-\ln p_A^{({\rm bulk})} \label{eq:end2}\\
&=&\frac{\ln 2}{8\pi G_NR_{\rm AdS}}{\rm Area}[\Gamma_A^0]+\frac{\ln 2}{\pi\hbar}I[\Gamma_A^0]\;.
\end{eqnarray}
The conjectured part (\ref{eq:hol}), that is, an equality between entropy (i.e., lost information) in the boundary and the bulk is based on the holographic principle.
Here, the integral measure ${\cal D}^f\Gamma_A$ satisfies
\begin{equation}
\int 1{\cal D}^f\Gamma_A=W_A^{({\rm bulk})}\;.
\end{equation}

% ----------------------------------------------------------
% An example of estimation of the action
% ----------------------------------------------------------

\smallskip
\smallskip

{\it Hydrodynamics.---}Next, to estimate the abbreviated action $I_A$, let us consider the hydrodynamics of the CFT$_2$ of the boundary (i.e., the system $A$ plus the reservoir), which is perfectly described by conserved quantities and does not require the local equilibrium condition because it is two-dimensional.\cite{D2,Review1,Review2}
Specifically, we consider the gravity dual of a one-dimensional perfect fluid with uniform and steady fluid two-velocity $u^x(t,x)=u$ and $u^0=\gamma$ such that
\begin{equation}
u_\mu=\eta_{\mu\nu}u^\nu\;,\ \ u^\mu u_\mu=-1\label{eq:u}
\end{equation}
for the Lorentz metric $\eta_{\mu\nu}$ of the two-dimensional boundary space-time, energy density $\varepsilon$, pressure $p=\varepsilon$, and `{\it mass}' density $\rho\equiv \varepsilon/\gamma^2$; we denote the kinetic part of $\varepsilon$ by $\varepsilon_{\rm kin}$.
Then, the metric of the bulk space-time, that is, the Schwarzschild AdS$_3$ black hole (i.e., the nonrotating Ba$\tilde{{\rm n}}$ados--Teitelboim--Zanelli black hole\cite{BTZ}), Lorentz boosted in the bulk space-time with the two-velocity $u^\mu$, is
\begin{equation}
ds^2=-2u_\mu dx^\mu dr+r^2\biggl[\eta_{\mu\nu}+\frac{r_+^2}{r^2}u_\mu u_\nu\biggr]dx^\mu dx^\nu\;.\label{eq:ds}
\end{equation}
In this metric, $x^\mu$ is covariantized with respect to Lorentz transformations in the boundary directions, where the index of $dx^\mu$ is lowered by $\eta_{\mu\nu}$ due to (\ref{eq:u}).
Here, we scale the curvature radius of AdS$_3$ space-time to unity (i.e., $R_{\rm AdS}=1$) and denote the radius of the event horizon of the Schwarzschild AdS$_3$ black hole taken to be equal to unity by $r_+$\cite{Complexity2,Complexity3}.
We denote the discrepancy between the metrics $\gamma_{\mu\nu}$ and $r^2\eta_{\mu\nu}$ by $\delta g_{\mu\nu}\equiv \gamma_{\mu\nu}-r^2\eta_{\mu\nu}$.
Then, from the one-dimensional Stefan--Boltzmann law of fluid of massless particles, we obtain $8\pi G_N p^x=-\delta g^{0x}=-r_+^2 \gamma u=-\varepsilon u/\gamma=-\rho \gamma u$ for the horizontal bulk-space momentum density $p^x$.\cite{AdSCFT3}
In addition to this, we have $v_x=-u$.

We let the domain of the system in the boundary be $A:[0,l]$.
Because (\ref{eq:ds}) is three-dimensional, (\ref{eq:ds}) is locally the metric of AdS$_3$\cite{Compere} and so is compatible with Swingle's interpretation of AdS/CFT correspondence; this fact is why we set this study in the context of AdS$_3$/CFT$_2$.
In the Poincar${\acute{{\rm e}}}$ coordinates, the geodesic $\gamma_A^0$ in the bulk is a half-circle.
The area element in the bulk space is given in the Poincar${\acute{{\rm e}}}$ metric of AdS$_3$ by
\begin{equation}
dA=\frac{dwdz}{z^2}\;,
\end{equation}
where $z=0$ represents the boundary.
Using these, in the `{\it non-relativistic}' regime (i.e., $v_x\ll 1$), when the lattice constant $\epsilon$ is sufficiently small relative to $l$ and $e^{r_+l}\gg 1$, we obtain the abbreviated action
\begin{eqnarray}
\frac{1}{\pi \hbar}I_A&=&\biggl[\frac{1}{8\pi G_N}\int_{\Gamma_A^0}dA\biggr]\frac{\rho u^2}{\pi \hbar}\\
&\simeq&\biggl[\frac{1}{8\pi G_N}\frac{l}{\epsilon}\biggr]\frac{4\varepsilon_{\rm kin}}{h}\\
&=&\biggl[\frac{c}{12\pi}\frac{l}{\epsilon}\biggr]t_\perp^{-1}\;.
\end{eqnarray}
Here, 
\begin{equation}
t_\perp=\frac{h}{4\varepsilon_{\rm kin}}\label{eq:ML}
\end{equation}
is the minimum elapsed time (the {\it Margolus--Levitin time}) required to execute a binary classical logic gate as a change of an original quantum pure state to another quantum pure state orthogonal to the original state (that is, a classical mechanically different state)\cite{ML,Lloyd,Simple}, and $c=3/2G_N$ is the central charge of the CFT$_2$ of the boundary\cite{BH}.

With this hydrodynamics, the proposal (\ref{eq:result}) is reduced to two conceptual arguments in terms of complexity:
i) after introducing quantum measurement errors, the quantum mixed state of the boundary system $A$, obtained by (\ref{eq:change}), has {\it classical probabilities} (i.e., probabilities with no interference).
In addition to this, $\Gamma_A^0$ is the holographic dual of this quantum mixed state of $A$ because the relative phases in $|\psi_A\ra$ are redundant for the geometry of $\Gamma_A^0$.
Then, $\Gamma_A^0$ is the spatial realization of the minimum program (i.e., the program with no redundancy), whose union of $N$ replicas outputs the statistical mixture (\ref{eq:change}) of $N$ product simultaneous eigenstates $\{|\{{\cal R}^\prime\}\ra\}$ of redefined positions $\{\widehat{Q}^\prime\}$ and redefined momenta $\{\widehat{P}^\prime\}$ of $A$, that is, the quantum mixed state of $A$ in the limit as $N$ goes to infinity.
This output is done independently by means of $\Gamma_A^0$ itself and a binary classical computer with the Margolus--Levitin time $t_\perp$.
The spatial length of this program is
\begin{equation}
\ell_A^s=\frac{1}{8\pi G_N}{\rm Area}[\Gamma_A^0]\;.\label{eq:ellAs}
\end{equation}
ii) The dimensionless length of this minimum program is
\begin{equation}
\ell_A=\ell_A^s+\frac{\ell_A^s}{t_\perp}\label{eq:ellA}
\end{equation}
and is the Shannon entropy (\ref{eq:HAbit}) of the quantum mixed state (\ref{eq:change}) of $A$.
If $t_\perp$ were not the shortest Margolus--Levitin time, then $\ell_A^s$ would not be minimal because the true length of the minimum program is unique.
In these arguments, whereas $\ell_A^s$ is independent of the state of fluid, $t_\perp$ is determined by the state of fluid.
Argument ii) is based on the fact that the Shannon entropy of a quantum mixed state can be written as the ensemble average of the classical Kolmogorov complexity.\cite{Equality}\footnote{See (\ref{eq:Hol}) in the appendix.}

\smallskip
\smallskip

{\it Consideration of the proposal in two models.---}To now, we have studied the holographic Shannon entropy of the CFT$_2$ coherence in the bulk.
There, we relied on classical statistical mechanics and invoked the ideas of complexity.
In this section, considering two one-dimensional models, we check our proposal (\ref{eq:result}) for the boundary qubits state by using a discrete holographic tensor network.
Here, the Shannon entropy of the coherence of the target system $A$ can be written in the Boltzmann form $H_A^{\rm bit}=(1/N)\log_2W_A^{({\rm bdy})}$ in terms of the number of classicalized (boundary) microstates $W_A^{({\rm bdy})}$ due to the measurement errors and the number $N(\gg 1)$ of elements in the statistical ensemble of $A$.

First, we consider the ground state of the target system $A$ in a strongly coupled CFT$_2$, where the classicalized microstates are counted by their real-space degrees of freedom (dual to the MERA network of $A$) only, due to $\epsilon_Q=\epsilon$.
We recall that, because of the Stirling formula, the disentangler operations and the isometries of EPR pairs in each layer of the MERA reduce the number of classicalized microstates to $1/2^N$ of the unreduced number per site in the deeper layer of the MERA.
In this way, we obtain $W_A^{(m)}/W_A^{(m+1)}=2^{Nl_{m+1}}$ for the number of classicalized microstates $W_A^{(m)}$ and the number of sites $l_m$ in the $m$-th layer ($m=0,1,\ldots,M$) of the MERA of $A$.
Here, the $0$-th layer is the boundary.
At the deepest $M$-th layer, we have $W_A^{(M)}=1$.
Consequently, we find that
\begin{eqnarray}
\frac{1}{N}\log_2 W_A^{(0)}&=&\frac{1}{N}\log_2\Biggl(W_A^{(M)}\prod_{m=0}^{M-1}\frac{W_A^{(m)}}{W_A^{(m+1)}}\Biggr)\\
&=&\sum_{m=1}^Ml_m\;,
\end{eqnarray}
which is the discretized area of $\Gamma_A^0$.
The generalization of this result to a finite temperature system is straightforward by applying the result of Ref.\cite{MIH}.

Second, we consider the thermal and momentum equilibrium state of the target system $A$ in a CFT$_2$ perfect fluid in the `{\it non-relativistic}' regime.
This momentum equilibrium state of $A$ is obtained from the thermal equilibrium state of $A$, dual to two MERA networks of $A$\cite{MIH}, by transforming the coordinate system $(t,x)$ of the boundary alongside the moving reservoir and transforming the Hamiltonian $\widehat{{\cal H}}_A$ of $A$, using
\begin{eqnarray}
t&\to&t\;,\label{eq:trf0}\\
x&\to&x-{\rm v}_xt\;,\label{eq:trf1}\\
\widehat{{\cal H}}_A&\to& \widehat{{\cal H}}_A-{\rm v}_x\widehat{{\cal P}}_A^{x}+{\cal K}_{{\rm v}_x}\label{eq:trf}
\end{eqnarray}
for the total momentum $\widehat{{\cal P}}_A^{x}$ of $A$ and the kinetic energy ${\cal K}_{{\rm v}_x}$ of $A$ with the velocity ${\rm v}_x$ of the reservoir.\cite{McLennan}
Here, we regard the average of the second term in (\ref{eq:trf}) with respect to $A$ as the Hamiltonian of the {\it a posteriori} momentum constraint $C$ of $A$ with the multiplier $-{\rm v}_x$ of $C$ and decompose this constraint $C$ into its minimal (mutually exclusive) momentum subconstraints.
Then, because the phase-space description of $A$ is well-defined by the classicalization due to the measurement errors, when we count the number of classicalized microstates of $A$, there are as many replicas of the MERA of $A$ (in addition to the original MERA of $A$) in the real space as these minimal momentum subconstraints of $C$.
At the decomposition of the constraint $C$, note that the energy $\varepsilon_{\rm min}=h/4R_{\rm AdS}$ has the thermal time (i.e., the decoherence time) $R_{\rm AdS}$\cite{Complexity2,Complexity3} as its Margolus--Levitin time.
Namely, $\varepsilon_{\rm min}$ is the minimum accessible energy width defined for a subconstraint of $C$.
Then, the number of minimal momentum subconstraints of $C$ (i.e., the number of replicas of the MERA of $A$ in the real space) is the division of the energy of $C$ by this energy width $\varepsilon_{\rm min}$, that is, $R_{\rm AdS}/t_\perp$ for the Margolus--Levitin time $t_\perp$ given by (\ref{eq:ML}).
Now, the number of classicalized microstates $W_A^{({\rm bdy})}$ is the product of the numbers of classicalized microstates $W_A^{(0),i}$ of the $i$-th MERA of $A$
($i=0,1,\ldots,R_{\rm AdS}/t_\perp$).
That is, we have
\begin{eqnarray}
W_A^{({\rm bdy})}=W_A^{(0),0}\times W_A^{(0),1}\times \cdots \times W_A^{(0),R_{\rm AdS}/t_\perp}\;.
\end{eqnarray}
The number of classicalized microstates of the MERA of $A$ is
\begin{equation}
W_A^{(0),i}=W_A^{(0)}\;,\ \ i=0,1,\ldots,\frac{R_{\rm AdS}}{t_\perp}\;.
\end{equation}
Consequently, we find that
\begin{equation}
W_A^{({\rm bdy})}=W_A^{(0)}\times \bigl(W_A^{(0)}\bigr)^{R_{\rm AdS}/t_\perp}\;.
\end{equation}
Namely, in this case, the Shannon entropy is the sum of the discretized area of the bulk domain $\Gamma_A^0$ and an abbreviated action term.
This is the content of our proposal (\ref{eq:result}).

\smallskip
\smallskip

{\it Conclusion: the holographic tensor network action.---}Finally, we derive the fundamental consequence of our proposal (\ref{eq:result}) for the AdS$_3$/CFT$_2$ correspondence by using two relations.

The first relation is the equality between the Shannon entropy in nats $H_A^{\rm nat}$ of the coherence of the thermofield-double pure state of the target system $A$ at thermal equilibrium and the quantum thermodynamic entropy $S_A^{\rm eq}$ of $A$ defined as the von Neumann entropy of the canonical distribution of $A$ after the redefinition of the phase-space variables.
That is, we have
\begin{eqnarray}
H_A^{\rm nat}&=&S_A^{\rm eq}\label{eq:JSU}\\
&=&\beta (E_A^{\rm eq}- F_A^{\rm eq})
\end{eqnarray}
for the inverse temperature $\beta$, the energy $E_A^{\rm eq}$, and the Helmholtz free energy
\begin{equation}
F_A^{\rm eq}\equiv -\frac{1}{\beta}\ln {\rm tr}e^{-\beta \widehat{{\cal H}}_A^\prime}\label{eq:FAdef}
\end{equation}
of $A$  as manyat the thermal equilibrium.
In (\ref{eq:FAdef}), the operator $e^{-\beta \widehat{{\cal H}}_A^\prime}$ for the redefined Hamiltonian $\widehat{{\cal H}}_A^\prime$ of $A$ can be written as
\begin{equation}
e^{-\beta \widehat{{\cal H}}_A^\prime}=\sum_k e^{-\beta \la E\ra_k}|\{{\cal R}^\prime\}_k\ra\la \{{\cal R}^\prime\}_k|\;,
\end{equation}
where
\begin{eqnarray}
\widehat{{\cal H}}_A&=&\sum_n E_n|E_n\ra\la E_n|\;,\label{eq:HA1}\\
\widehat{{\cal H}}_A^\prime&=&\sum_nE_n\Biggl(\sum_k p_{n,k}|\{{\cal R}^\prime\}_k \ra\la \{{\cal R}^\prime\}_k|\Biggr)\label{eq:HA2}\\
&=&\sum_k\la E\ra_k|\{{\cal R}^\prime\}_k\ra\la \{{\cal R}^\prime\}_k|\;.\label{eq:HA3}
\end{eqnarray}
In (\ref{eq:HA1}) and (\ref{eq:HA2}), we assume the conditions $\la E_m|E_n\ra=\delta_{mn}$ and $\la \{{\cal R}^\prime\}_k|\{{\cal R}^\prime\}_{k^\prime}\ra=\delta_{kk^\prime}$, respectively.
Then, we have the relations $\sum_k p_{n,k}=1$ and $\la E\ra_k=\sum_n E_np_{n,k}$.
This equality (\ref{eq:JSU}) is valid under replacement of {\it thermal equilibrium} by {\it thermal and momentum equilibrium}.

The second relation is the mainstay relation of the AdS$_3$/CFT$_2$ correspondence in the equilibrium case\cite{GKP,W,AdSCFT2},
\begin{equation}
\beta F_{{\rm CFT}_2}^{\rm eq}=\frac{1}{\hbar}I_{{\rm AdS}_3}\;,\label{eq:GKPW}
\end{equation}
for the equilibrium Helmholtz free energy $F_{{\rm CFT}_2}^{\rm eq}$ of CFT$_2$ in the large-$N$ limit\cite{tHooft} and the classical gravity action $I_{{\rm AdS}_3}$ evaluated on AdS$_3$.

Combining (\ref{eq:result}), (\ref{eq:JSU}), and (\ref{eq:GKPW}), we conjecture that gravity in AdS$_3$ can be replaced by the system with the action
\begin{equation}
I_{\rm bulk}=-\frac{b}{\pi}W_{\rm TN}-\hbar b A_{\rm TN}\;,\ \ b=\ln 2\label{eq:ITN}
\end{equation}
as the bulk theory equivalent to the boundary CFT$_2$ at thermal and momentum equilibrium.
Here, $W_{\rm TN}$ denotes our abbreviated action $I_A$, where $A$ refers to the holographic tensor network; $A_{\rm TN}$ denotes the discretized area of the holographic tensor network; and $b$ is the information of a classical bit and indicates that the holographic tensor network consists of qubits.

We can observe two important points from (\ref{eq:ITN}).

First, the first term of (\ref{eq:ITN}) is the abbreviated action defined for the `{\it temporal}' extra dimension $r$.
From the abbreviated principle of stationary action subject to the CFT$_2$ boundary condition and energy conservation, this result suggests that, in the realized holographic tensor network, the horizontal information propagation is optimal with respect to the `{\it time}'.
This is in agreement with (\ref{eq:ds2}).

Second, the second term of (\ref{eq:ITN}) takes the form of the action of a membrane with the tension
\begin{equation}
{\cal T}=\frac{\hbar b}{8\pi G_NR_{\rm AdS}^2}\;,
\end{equation}
which is proportional to the Planck constant.
This result suggests that the bulk space is quantized by the energy ${\cal T}$ per unit bulk space area.

We conclude this paper with the proposal (\ref{eq:result}) and this change of the perspective (\ref{eq:ITN}) of the AdS$_3$/CFT$_2$ correspondence.

\begin{appendix}

\section{Determination of equal a priori probability $p_A^{({\rm bulk})}$}

In this appendix, we determine the {\it equal a priori probability} $p_A^{({\rm bulk})}$ for a given metric of the bulk space-time.

After introducing quantum measurement errors, we consider $n_r(\gg 1)$ replicas of the $\mu$-space ${\mathfrak M}_r$ of a site of the holographic tensor network and superpose these replicas; ${\mathfrak M}_r$ is defined at each position $r$ in the extra dimension.
We denote by $n_r(x,p^x)$ the occupied number density at the point $(x,p^x)$ in the superposed ${\mathfrak M}_r$.

The fundamental postulate of statistical mechanics asserts that the ensemble of equally likely microstates subject to constraints is in a state of statistical equilibrium.
So, the problem is subject to the constraints
\begin{align}
n_r&=\iint n_r(x,p^x) dxdp^x\label{eq:Constr1}\\
&={\rm a\ large\ constant}\;,\\
p_r^x(x)&=\int p^x n_r(x,p^x)dp^x\label{eq:Constr2}\\
&={\rm a\ posteriori\ constants}\;,
\end{align}
to determine the configuration $n_r(x,p^x)$ that maximizes the number of microstates corresponding to the macrostate specified by (\ref{eq:Constr1}) and (\ref{eq:Constr2})
\begin{equation}
W_r=\frac{n_r!}{\prod_{x}\prod_{p^x}n_r(x,p^x)!}\;.
\end{equation}
For $p^x_r(x)$, we suppose only the conditions that maintain their values while we maximize the number of microstates, and the values of $p_r^x(x)$ would be determined {\it a posteriori} after the distribution $n_r(x,p^x)$ is determined.
Here, due to
\begin{equation}
\ln W_r=n_r\ln n_r-\iint n_r(x,p^x)\ln n_r(x,p^x) dx dp^x\;,
\end{equation}
the variational equations are
\begin{align}
0&=\delta \ln W_r\\
 &=-\iint (\ln n_r(x,p^x)+1)\delta n_r(x,p^x)dx dp^x\;,\\
0&=\iint \delta n_r(x,p^x) dxdp^x\;,\\
0&=\int p^x \delta n_r(x,p^x)dp^x\;.
\end{align}

For consistency with the thermodynamic relation 
\begin{equation}
\frac{\pa \Delta s_A^{({\rm bulk})}}{\pa p^x}=-v_x\Delta \biggl(\frac{1}{T_{\rm bulk}}\biggr)
\end{equation}
under the identification (\ref{eq:id}) in the main text, we introduce the Lagrange multipliers for the constraints as
\begin{equation}
0=\iint (\ln n_r(x,p^x)+\alpha +\beta p^xv_x)\delta n_r(x,p^x)dxdp^x\;,\label{eq:TheEq}
\end{equation}
where the multiplication by unity with two spatial dimensions is omitted for the last two terms.
Then, we obtain
\begin{equation}
n_r(x,p^x)\propto e^{-\alpha-\beta p^xv_x}\;,
\end{equation}
where $n_r$ in (\ref{eq:Constr1}) diverges and so, it is an {\it a posteriori} constant.
From this, we have
\begin{align}
n_{\Gamma_A}[p^x]&\equiv\prod_{(x,r)\in \Gamma_A}n_r(x,p^x)\\
&\propto\exp\biggl(-\alpha {\rm Area}[\Gamma_A]-\beta \int_{\Gamma_A}p^x v_x d\Gamma_A \biggr)\;.
\end{align}
This takes the form of
\begin{equation}
n_{\Gamma_A}[p^x]\propto e^{-\alpha {\rm Area}[\Gamma_A]-\beta I[\Gamma_A]}\;.\label{eq:nAt}
\end{equation}

Based on this result (\ref{eq:nAt}), by following the arguments from (\ref{eq:start}) to (\ref{eq:end2}) in the main text, we obtain
\begin{align}
S_A&=(\ln 2)H_A^{\rm bit}\\
&=\alpha {\rm Area}[\Gamma_A^0]+\beta I[\Gamma_A^0]\;,\label{eq:App}
\end{align}
where $I[\Gamma_A^0]$ is defined for the metric of the bulk space-time.

Now, because $H_A^{\rm bit}$ is the Shannon entropy of the statistical mixture (\ref{eq:change}) of the product simultaneous eigenstates $\{|\{{\cal R}^\prime\}\ra\}$ of redefined positions $\{\widehat{Q}^\prime\}$ and redefined momenta $\{\widehat{P}^\prime\}$ in bits, we obtain\cite{Equality}
\begin{equation}
H_A^{\rm bit}=\lim_{N\to \infty}\frac{K_A^{\rm bit}((|\{{{{\cal R}}}^\prime\}\ra)_{n\cdot N})}{N}\;.\label{eq:Hol}
\end{equation}
Here, the product simultaneous eigenvectors $|\{{{\cal R}}^\prime\}_a\ra$ of $\{\widehat{Q}^\prime\}$ and $\{\widehat{P}^\prime\}$ are encoded by length-$n$ binary sequences.
On the right-hand side of (\ref{eq:Hol}), these binary sequences belong to an infinitely large binary sequence, which represents the statistical mixture (\ref{eq:change}) of $N$ replicas of the system $A$, with length $n\cdot N$ in the limit as $N$ goes to infinity, and $K_A^{\rm bit}$ is the classical Kolmogorov complexity in bits.

From (\ref{eq:Hol}), we determine the two Lagrange multipliers $\alpha$ and $\beta$ by considering two cases.

First, we consider the case of a static bulk space-time; that is, we consider a system $A$ not in a state of momentum equilibrium.
In this case, the second term of (\ref{eq:App}) vanishes, and we assert that $H_A^{\rm bit}=(\alpha/\ln 2){\rm Area}[\Gamma_A^0]$ matches the holographic complexity $C_A$\cite{Complexity1,Complexity2,Complexity3,Alishahiha} because of (\ref{eq:Hol}), the reduction of $|\{{\cal R}^\prime\}\ra$ to $|\{Q^\prime\}\ra$, footnote 3, and the sufficient fineness of the introduced measurement errors $\sigma_Q$ and $\sigma_P$ (i.e., there is no extra loss of information).
From the expression of $C_A$ in Ref.\cite{Alishahiha}, $\alpha$ is determined by
\begin{equation}
\alpha=\frac{\ln 2}{8\pi G_NR_{\rm AdS}}\;.\label{eq:alpha}
\end{equation}

Second, we consider the case of a non-static bulk space-time; that is, we consider a system $A$ in equilibrium with a momentum reservoir having a non-zero velocity $u$.
From (\ref{eq:App}) and (\ref{eq:Hol}), we obtain
\begin{equation}
\lim_{N\to \infty}\frac{K_A^{{\rm bit}}((|\{{{\cal R}}^\prime\}\ra)_{n \cdot N})}{N}=\frac{\alpha}{\ln 2}{\rm Area}[\Gamma_A^0]+\frac{\beta}{\ln 2}I[\Gamma_A^0]\;.\label{eq:KA}
\end{equation}

When the phase $e^{iI_A/\hbar}$ advances from $1$ to $-1$ with $C_A$ unchanged, this advance of the phase increases (\ref{eq:KA}) by one.
Namely, in the ensemble average, this advance of the phase adds a binary classical logic gate to the minimum program of the quantum mixed state (\ref{eq:change}) of $A$ for a binary classical computer with the Margolus--Levitin time $t_\perp$, and the initial state of this minimum program has complexity $C_A$.
This argument is based on original arguments by Lloyd\cite{ML,Lloyd} and Brown {\it et al.} made within a context different from ours\cite{Susskind,Susskind2}.
From this argument and (\ref{eq:KA}), $\beta$ is determined by
\begin{equation}
\beta=\frac{\ln 2}{\pi\hbar}\;.\label{eq:beta}
\end{equation}

From (\ref{eq:App}), (\ref{eq:alpha}), and (\ref{eq:beta}), the {\it equal a priori probability} $p_A^{({\rm bulk})}$ is determined by (\ref{eq:pAbulk}) in the main text.

\end{appendix}

\end{document}